\documentclass[12pt]{article}
\usepackage{epsfig}

\date{}

\newcommand{\be}{\begin{eqnarray}}
\newcommand{\ee}{\end{eqnarray}}
\newcommand{\non}{\nonumber\\}

\title{\begin{flushright}\normalsize BNL-NT-00/5\\
LBNL-45248
\end{flushright}
\vskip 0.1in
Wong's equations and the small $x$ effective action in QCD}
\author{Jamal Jalilian-Marian\\
{\small \it Physics Department, University of Arizona, Tucson, AZ 85721.}\\
Sangyong Jeon\\
{\small \it Nuclear Science Division,}\\
{\small \it Lawrence Berkeley National Laboratory, Berkeley, CA 94720.}\\
Raju Venugopalan\\
{\small \it Physics Department, 
Brookhaven National Laboratory, Upton, NY 11973.}\\
}

\begin{document}
\bibliographystyle{unsrt}
\maketitle
                           
\begin{center}
{\bf Abstract}\\
\end{center}

\noindent We propose a new form for the small x effective action in 
QCD. This form of the effective action is motivated by Wong's equations 
for classical, colored particles in non--Abelian background fields. 
We show that the BFKL equation, which sums leading logarithms in x, is 
efficiently reproduced with this form of the action. We argue that 
this form of the action may be particularly useful in computing next-to-
leading-order results in QCD at small x. 

\vfill \eject

\baselineskip=22pt plus 1pt minus 1pt
\parindent=25pt

\section{Introduction}
\vskip 0.1in

One of the more interesting open questions in QCD is the behavior of 
cross--sections at very high energies~\cite{Mueller}. 
In the last decade, a kinematic 
window has opened up at colliders where $Q^2 \gg \Lambda_{QCD}^2$ but 
$x=Q^2/s \ll 1$. The physics in this regime is non--perturbative because  
the field strengths at small $x$ are large. However, it is also weak 
coupling physics since $\alpha_S(Q^2)\ll 1$. Further, since the density of 
partons is large at small $x$, classical field 
methods are applicable~\cite{Zakopane99}.

An effective field theory approach can be used to study the physics of 
small $x$ modes in 
QCD~\cite{MV,AJMV,RGE}~\footnote{For alternative approaches, 
see for example, Ref.~\cite{Lipatov2} and references therein, 
and Ref.~\cite{Kovchegov} 
and references therein. These will not be discussed in this paper.} . 
The small $x$ effective action 
is obtained by successively integrating out the more static modes at larger 
values of $x$. The measure for this action is represented by a weight 
corresponding to the color charge density of the higher $x$ modes. As 
one integrates out higher $x$ modes, the form of the action is maintained, 
while the weight satisfies a Wilsonian non--linear renormalization 
group (RG) equation~\cite{RGE}. If the parton density is not too large, the 
RG equation can be linearized, and the resulting equation is the 
well known BFKL equation. The BFKL equation is a renormalization 
group equation that 
sums the leading logarithms in $\alpha_S\ln(1/x)$~\cite{Lipatov}.
In the double 
log limit of small $x$ and large $Q^2$, the Wilson RG can be simplified, and 
one obtains a series in inverse powers of $Q^2$, where the 
leading term is 
the small $x$ DGLAP equation~\cite{DGLAP} and the first sub--leading 
term agrees with the expression derived by Gribov, Levin and 
Ryskin~\cite{GLR}, and by Mueller and Qiu~\cite{MuellQiu}. 

The effective action approach therefore reproduces the standard linear
evolution equations of perturbative QCD in the limit of low parton
densities. The truly interesting and unknown regime however is the
non--linear regime of high parton densities where one might hope to
predict novel phenomena~\cite{Levin,AlMueller,FrankStrik}. What the
correct effective action is in the high density regime should
therefore be a matter of some interest.

In this paper, we will discuss an alternative gauge invariant form to
the gauge invariant action discussed in Ref.~\cite{RGE}.  The
motivation for this form of the effective action came from our recent
work in formulating a many body world line formalism for the one loop
effective action in QCD~\cite{JSRJ}. Briefly, the difference between
the two actions is in the term describing the coupling of the small
$x$ gauge field modes to the large $x$ modes represented by a color
charge density $\rho$. In the work of Jalilian--Marian, Kovner,
Leonidov, and Weigert (JKLW), this term is expressed as
\be
S_{int}^{JKLW}\sim {\rm Tr} \left(\rho\,W_{\infty,-\infty}\right)\, , \nonumber
\ee
where $W$ is an adjoint matrix corresponding to a path ordered 
exponential of the gauge field $A^-$ in 
the light cone direction $x^+$. We propose instead that this term be 
\be
S_{int}\sim {\rm Tr}\left(\rho\,\ln W_{\infty,-\infty}\right)\, ,\nonumber
\ee
replacing $W\rightarrow \ln W$ in the effective action.

The earlier form of $S_{int}$ was chosen primarily because it is a 
gauge invariant generalization of the coupling between hard and soft 
modes. The motivation for the latter form comes from the background 
field method and the eikonal approximation. The one loop effective action, 
in the background field method, can be expressed as 
 $\ln[{\rm det}(D^2)]\rightarrow {\rm Tr}
\ln[D^2]$, where $D$ is the usual covariant derivative. If, for instance, 
one integrated out hard fermions in the soft background gauge 
field, the eigenvalues of the determinant 
would correspond to solutions of the Dirac equation in the eikonal 
approximation. These correspond to path ordered phases of the 
component of the soft gauge field, conjugate to the hard current, in the 
fundamental representation~\cite{Nachtmann91}. Similarly, performing 
an eikonal separation of hard and soft gauge fields, one obtains path 
ordered exponentials (in the adjoint representation) 
of the soft gauge fields (see, for example, appendix B of 
the first paper in Ref.~\cite{RGE}). Since the 
effective action is the logarithm of the determinant, one can thus 
anticipate the appearance of 
the logarithm of the path ordered phase in the effective action. This 
form of the effective action is also gauge invariant. We will show 
later that the $\ln(W)$ action has the nice feature that one can 
derive the BFKL equation from it efficiently--certain terms that one needs 
to argue to be zero in the $W$ form of the effective action are absent 
in the $\ln(W)$ action.

The subsequent discussion is organized as follows. In section 2, we will 
discuss the form of the small $x$ effective action discussed in 
Ref.~\cite{RGE}. In section 3, we will discuss Wong's equations and 
motivate an alternative form for the small $x$ effective action. We will 
show that the form of the action that we propose is also consistent with 
Wong's equations and that the two different currents arising from the
two actions correspond to different boundary conditions for solving 
Wong's equations. In section 4, we 
will show that our form of the 
effective action also reproduces the BFKL equation. We end this 
paper with a brief summary in section 5.
Some technical details are contained in three appendices.

\section{The Small $x$ Effective Action}
\vskip 0.1in

In this section, we will review the effective action and Wilson 
renormalization group approach to small $x$ QCD as developed 
in Refs.~\cite{MV,AJMV,RGE}. We refer the reader to these papers for more 
details. 

We start with the following action~\cite{RGE} which is the gauge
invariant generalization of McLerran-Venugopalan effective action first 
proposed in \cite{MV}. In the infinite momentum frame, and in Light 
Cone gauge $A^+=0$, one can write 
\be
S 
&=& -{1\over 4} \int d^4 x G_a^{\mu\nu}G^a_{\mu\nu}  
+ i \int d^2 x_\perp F[\rho ^a(x_\perp)] 
\non
&+& 
{{i}\over{N_c}} \int d^2 x_\perp dx^-
\delta (x^-)
\,{\rm Tr}\left(\rho (x_\perp) 
W_{\infty,-\infty} [A^-](x^-,x_\perp)\right)\,, 
\label{eq:action}
\ee
where $W$ is the Wilson line in the adjoint representation along the
$x^+$ axis
\be
W_{\infty,-\infty}[A^-](x^-,x_\perp) = 
\hat P\exp 
\bigg[ig \int_{-\infty}^\infty dx^+ A^-_a(x^+,x^-,x_\perp)\,T_a \bigg]. 
\label{Wilsonline}
\ee
and the label $\hat P$ denotes the path-ordered exponential.

Taking the saddle point of the effective action, we obtain the Yang--Mills 
equations 
\be
D_\mu G^{\mu\nu}_{a}= \delta^{\nu +} J^{+}_{a}\,,
\label{eq:eom1}
\ee
with the current 
\be
J^+_a(x)={g\over{N_c}} 
\delta (x^-) 
{\rm Tr}\Bigg[T_a W_{x^+, -\infty} [A^-] 
\rho (x_\perp ) W_{\infty ,x^+} [A^-]\Bigg]\, 
\label{eq:current}
\ee
satisfying the boundary condition
\be
J^+_a(x^+=-\infty)={g\over{N_c}} 
\delta (x^-)  
{\rm Tr}\Bigg[T_a \rho (x_\perp )  W_{\infty,-\infty} [A^-] 
\Bigg]\, 
\label{eq:bc}
\ee
The first term in the expansion of the Wilson line in the action is 
\be
S_{\rm int}= -g\,\int d^4 x\, A^-\rho(x_\perp)\delta(x^-)
\label{eq:naiveaction}
\ee 
used in \cite{MV}.

To derive the general evolution equation, one first solves the classical 
equations of motion, computes quantum fluctuations 
in the background of the classical field (semi-classical approximation), 
and separates these fluctuations according to their 
longitudinal momentum as 
\be
A^a_{\mu} (x) = b^a_{\mu} (x) + \delta A^a_{\mu} (x) + a^a_{\mu} (x)\,,
\label{eq:decomp}
\ee
where $ b^a_{\mu} (x)$ is the solution of the classical equations of
motion, $ \delta A^a_{\mu} (x) $ is the fluctuation field containing 
longitudinal momentum modes $k^+$  that are constrained to be 
$p^+< k^+< P^+$. 
The upper cutoff
$P^+$ is the longitudinal momentum of the fast moving
charges while the lower cutoff $p^+$ is the 
momentum scale of the soft fluctuations. These cut-offs are chosen to 
be such that $\alpha_S\ln(P^+/p^+)\ll 1$ since quantum fluctuations give 
rise to such logarithms~\cite{AJMV}. This constraint thus requires  
that the fluctuations with momentum modes $p^+< k^+< P^+$ are small, 
and can therefore be integrated out to obtain the 
effective action for the soft (in longitudinal momenta alone!) fields 
$a^\mu$. This procedure can be iterated as one goes to smaller $x$ leading 
to a Wilsonian RG equation~\cite{RGE}.

The physics underlying this procedure is simple. One starts with some
initial color charge density at large $x$ represented by $\rho$. In
order to compute a quantity with this action, one averages over all
color configurations represented by the statistical weight
\be
 Z = \exp\{-F[\rho]\} \, . \nonumber
\ee
We then integrate out the hard fluctuations with the constraint
discussed above. This changes the color charge density and the
statistical weight for their configurations.  The soft fluctuations,
with logarithmic accuracy, ``see'' the  induced charge density as a
part of the color charge density to which they are coupled. As one goes
to smaller and smaller $x$ (longer and longer wavelength gluons) one
correspondingly includes more of the hard fluctuations in the color
charge density. One obtains the following renormalization group
equation for the change of the statistical weight $Z$ with
$x$~\cite{RGE}:
\be
{d Z\over {d \ln(1/x)}}=\alpha_S \left[{1\over 2} {\delta^2\over
{\delta \rho_\mu \delta \rho_\nu}}\left(Z\chi_{\mu\nu}\right)
-{\delta\over
{\delta\rho_\mu}} \left(Z\sigma_\mu\right)\right] \, ,
\label{nonlinRG}
\ee
where $\sigma[\rho]$ and $\chi[\rho]$ are respectively one and two
point functions obtained by integrating over $\delta A$ for fixed
$\rho$. The one point function $\sigma$ includes the virtual
corrections to $F[\rho]$ while the two point function $\chi$ includes
the real contributions to $F[\rho]$. Both of these can be computed
explicitly from the small fluctuations propagator in the classical
background field. In the weak field limit, the functions $\sigma$ and 
$\chi$ simplify, and the resulting renormalization group equation is 
the BFKL equation.

In the following section, we will interpret the color charge density
$\rho$ of the hard (large $k^+$) modes as the density of classical color
charges moving in the field of the soft modes. Such an
interpretation arises naturally when one computes the one loop
effective action in QCD for soft modes using the background field
method~\cite{JSRJ,Strassler,Pisarski97}. One expects therefore
that these classical charges must satisfy Wong's equation for the motion 
of color charges in a non--Abelian background field. These equations are
discussed below where a new form of the effective action is
proposed. 

\section{Wong's equations and an alternative effective action}
\label{sec:logwaction}
\vskip 0.1in

In Ref.~\cite{JSRJ}, we developed a many body formalism for the one loop 
effective action in QCD. We employed the world line 
formalism~\cite{Strassler,dHokergagne}
to re--write the path ordered exponential as a quantum mechanical 
path integral over world lines. The equations of motion for the 
corresponding point particle Lagrangian satisfies 
Wong's equations for the motion of a classical charged particle in a 
non--Abelian background field~\cite{SKWong}. These are 
\be
p^\mu & = & m {dx^\mu\over d\tau} = mv^\mu\\
{dp^\mu\over d\tau}
& = &
v_\nu\, Q^a\, G^{\mu\nu}_a\\
D^\nu G_{\nu\mu} & = & j_\mu
\label{SKW}
\ee
where
\be
j_\mu(x)= \int d\tau\, Q(\tau)\, v_\mu(\tau)\, 
\delta^{4}\left[x - z(\tau)\right].
\label{eq:color_j}
\ee
and
\be
\dot{Q}= -ig\left[Q, v_\mu A^\mu\right]
\label{eq:charge}
\ee
The generalization to a system of particles is straightforward.

Without explicitly going over to the world line approach, one can write down 
the following many-body classical action
 \be
 S_{\rm Wong} 
 =  
 - 
 {1\over 4}
 \int d^4x\,
 G_{\mu\nu}^a G^{\mu\nu}_a
 -\sum_{I=1}^K\,
 \int d\tau\,
 m_0^I\,\sqrt{v_\mu^I v_I^\mu} 
 +
 {i\over N_c}\, \sum_{I=1}^K{\rm Tr}\left\{Q_I\, \ln W_I\right\}\,,
 \non
 \label{jsrj1}
 \ee
 where $K$ is the number of Wong's particles and $I$ is the particle
 label. Also 
 \be
 W_I
 =
 \hat{P} \exp\left(ig\int_{-\infty}^\infty\,d\tau\,
 v_\mu^I\, A^{\mu}_a(x_I^\mu(\tau))\,T_a^I
 \right)\, .
 \label{eq:Wilson}
 \ee
 This action is gauge invariant under gauge transforms $U$ that satisfy
 the constraint  $U(\infty) = U(-\infty)$. We will define ``$\ln W$'' 
 shortly. As shown in appendix A, the 
 Wong equations in Eq.~(\ref{SKW}) can be derived from 
 Eq.~(\ref{jsrj1}) above.   

 In an infinite momentum frame (relevant for the small $x$ problem), 
 the momenta of the particles are not dynamical. They are static light 
 cone sources--$v^{\mu} = \delta^{\mu +}$. 
The kinetic part of the action in $S_{\rm Wong}$ therefore drops out 
 to yield
 \be
 S_{\rm Wong} 
 = 
 - 
 {1\over 4}
 \int d^4x\,
 G_{\mu\nu}^a G^{\mu\nu}_a
 + 
 {i\over N_c}\, \sum_{I=1}^K\, {\rm Tr}\left\{Q_I\, \ln W_I \right\}\, .
 \ee
 
 We assume now that the initial $x_I^- = 0$ is the same for all
 particles. 
 In the infinite momentum frame, $P^+\rightarrow \infty$, this 
 assumption is justified because the 
 particles can be viewed as being confined to a Lorentz contracted sheet 
 in the transverse plane of width $1/P^+$. 
 This implies that the particles can be labeled using  
 their transverse positions $x_\perp^I$ only.
 Using 
 \be
 \rho_a(x_\perp)
 =
 \sum_{I=1}^K
 \delta(x_\perp - x^I_\perp)
 Q^I_a\, ,
 \ee
 one can assume $\rho^a(x_\perp)$ to be continuous (and large). One can 
 therefore make an educated guess that
 the coarse grained effective action of the wee parton modes will be 
 \be
 S_{\ln W} =
 - 
 {1\over 4}
 \int d^4x\, G_{\mu\nu}^a G^{\mu\nu}_a
 + 
 {i\over N_c}\,\int d^2x_\perp\, {\rm Tr}\left\{
 \rho (x_\perp)\,\ln W(x_\perp) \right\}\non
 \label{logwaction} 
\ee
 where now
 \be
 W(x_\perp) =\hat{P}\exp\left(
 ig\int_{-\infty}^{\infty}dx^{+}\,
 A^{-}_a(x^+, 0, x_\perp)\,T_a\right)\, .
\label{poe}
 \ee
 Just as in Eq.~(\ref{eq:action}), the action $S_{\ln W}$ should
 contain an identical functional $F[\rho]$ representing the likelihood
 of different $\rho$ configurations. This term will only be implicit in 
 what follows since it is not relevant to the concerns of this 
 paper.

 We will now show explicitly that the charge obtained from the action 
 $S_{\ln W}$ is Hermitean and traceless, and therefore an element of 
 the Lie algebra.

We first define the log of an operator as the power series
\be
\ln W = \ln(1 - (1-W)) \equiv 
-\sum_{k=1}^\infty {1\over k} (1-W)^k.
\label{logdef}
\ee
Taking the functional derivative of $\ln W$ with respect to $A$ gives
\be
{\delta \over \delta A_\mu^a}\ln W= 
\sum_{k=1}^\infty {1\over k} \sum_{s=0}^{k-1} (1-W)^s 
{\delta W \over \delta A_\mu^a} (1-W)^{k-s-1}
\ee
After some straightforward algebra, this can be written as
\be
{\delta \over \delta A_\mu^a}\ln W = 
\int_0^1 d\lambda\,\, 
{1\over 1 - (1 - W)\lambda}\, 
{\delta W \over \delta A_\mu^a}\,\, 
{1\over 1 - (1 - W)\lambda} 
\ee
Then from the relation 
\be
J^{\mu}_{a}
=
-
{i\over N_c}\,
{\rm Tr}\left( \rho \, {\delta \over \delta A_\mu^a} 
\ln W\right)\, ,
\label{eq:wc}
\ee
we find that the color charge is given by 
\be
Q(x^+)
& = & \int\, d^3 x\, J^{+}(x)
\non
& = & 
\int\, d^3 x\, W(x^+, -\infty)\, 
\left[\int_0^1 d\lambda \,{1\over B(\lambda)}
\rho\, {1\over B(\lambda)} \right] W(\infty, x^+)
\label{eq:Qxplus}
\ee
where we used the shorthand
\be
B(\lambda) \equiv 1 - (1 - W)\lambda
\;
\ee
and
\be
W(x^+_f, x^+_i) =
\hat{P}
\exp\left(
ig\int_{x^+_i}^{x^+_f}dx^{+}\,
A^{-}_a(x^+, x^-, x_\perp)\,T_a
\right)
\;
\ee
We also defined $W \equiv W(\infty, -\infty)$.

It is easy to check that the current density $J^+$ satisfies 
\be
{\partial J^+\over \partial x^+} & = & -ig\left[J^+, A^-\right] \, ,
\label{wch}
\ee
and hence is a solution of the Wong's equation with a ``boundary''
condition given by
\be
J^+(x^+ = -\infty)= 
\int_0^1 d\lambda\, 
B^{-1}(\lambda)\,\rho \,B^{-1}(\lambda)  W
\label{eq:wbc}
\ee
Note that $1/B(\lambda) = B^{-1}(\lambda)$.

 To confirm that $J^+(x)$ is an element of the Lie-Algebra, first
 consider the trace.  We have
 \be
 {\rm Tr}\,\left(J^+(x)\right)
 & = &
 \rho^b\,{\rm Tr}\,\left(
 W\left[\int_0^1 d\lambda\left(B^{-1}(\lambda)\right)^2\,T_b\,\right]\right)
 \, ,
 \ee
 since $W$ and $B$ commute. 
 One can show that 
 \be
 {d\over d\lambda} B^{-1}(\lambda)
 =
 (-1)\,\left(B^{-1}(\lambda)\right)^2\,(W-1) \, .
 \ee
 Consequently, 
 \be
 {\rm Tr}\,\left(J^+(x)\right)
 & = &
 \rho^b
 {\rm Tr}\,\left(W\left[\int_0^1 d\lambda\,(B^{-1}(\lambda))^2\,
 T_b\,\right]\right) 
 \non
 & = & \rho^b\,{\rm Tr}\,\left(W\,(1-W)^{-1}\left(
 B^{-1}(1) - B^{-1}(0)\right)\,T_b\right)
 \non
 & = & \rho^b\,{\rm Tr}\,\left(W (1-W)^{-1}(W^{-1} - 1)\,T_b\right)
 \non
 & = & \rho^b\,{\rm Tr}\,\left(T_b\right) = 0.
 \ee

 We shall now show that $J$ is also Hermitean. Consider 
 \be
 (J^+(x))^\dagger & = & \rho^b\,\left(W(x^+,-\infty)\, 
 \left[\int_0^1 d\lambda\, B^{-1}(\lambda)\,T_b\,B^{-1}(\lambda)\right] 
 \,W(\infty, x^+)\,\right)^\dagger
 \non
 & = & \rho^b\, W(x^+,\infty)\,\left[\int_0^1 d\lambda\,
 (B^{-1}(\lambda))^\dagger\,T_b\, 
 (B^{-1}(\lambda))^\dagger\right]\,W(-\infty, x^+) 
 \non
 & = &
 \rho^b\, W(x^+,-\infty)\,\left[\int_0^1 d\lambda\,W^\dagger\,
(B^{-1}(\lambda))^\dagger\,T_b\,W^\dagger\,(B^{-1}(\lambda))^\dagger
 \right]W(\infty, x^+)\, .
\ee
 Let us now focus on the term in the square brackets. Since
 \be
 W^\dagger = W(-\infty, \infty) = W^{-1}\, ,
 \ee
 this term can be re-written as 
 \be
 \int_0^1 d\lambda\, (B^{-1}(\lambda)\,W)^\dagger\,T_b\, 
 (B^{-1}(\lambda)\,W)^\dagger \, .
 \ee
 Here one has used the relation $W\,B^{-1} = B^{-1}\,W$. Performing the change
 of variable 
 $\lambda\rightarrow 1-\lambda$, one can show that 
\be
(B^{-1}\,W) = (B^{-1})^\dagger \, .
\ee
Thus,
\be
(J^+(x))^\dagger &=&  \rho^b\, W(x^+,-\infty)\,
\int_0^1 d\lambda\,\left[(B^{-1}(\lambda)\,W)^\dagger\,T_b\, 
 (B^{-1}(\lambda)\,W)^\dagger  \right]W(\infty, x^+)\, ,\non
&=&  \rho^b\, W(x^+,-\infty)\,
\int_0^1 d\lambda\,\left[B^{-1}(\lambda)\,T_b\,B^{-1}(\lambda)
\right]W(\infty, x^+)\, ,\non
&=& J^+ (x) \, .
\ee
We have now explicitly shown above that $J^+$ (and hence $Q$) 
is both Hermitean and traceless.
It is therefore an element of the Lie Algebra. In general, it is possible, 
if non--trivial, to show that $\ln(W)$ itself is a member of the 
Lie algebra~\cite{HarryLam}.
The charge obtained from Eq.~(\ref{eq:current}) is also an element
of the Lie Algebra. It is easy to see that the color components
of the color charge $J^+_a$ are real and therefore, the color charge 
matrix defined as $J_{\mu}={1 \over N_c} J_{\mu}^a T^a$ is Hermitean 
and traceless.
Both $S_W$ and $S_{\ln W}$ lead to Wong's equations, but with
a different current $J_{\mu}$. This difference is
due to imposing different ``boundary'' conditions at $\tau = -\infty$
when solving the Wong's equations (\ref{eq:charge}) as given by
(\ref{eq:bc}) and ({\ref{eq:wbc}). It should be noted that the 
boundary condition
in ({\ref{eq:wbc}) is more complicated than (\ref{eq:bc}), and 
involves the non--trivial task of inverting the operator $B(\lambda)$.
It is important to realize that the two different currents may describe
different physics.

\section{The $\ln W$ action and the BFKL equation}
\vskip 0.1in

We will now show that the form of the action in Eq.~(\ref{logwaction}) also 
reproduces the BFKL equation. Since the two actions differ only by the 
form of the Wilson line term, we will focus on the expansion of the 
Wilson line term in the two actions. To reproduce the Wilsonian 
renormalization group evolution, we need to keep terms that are quadratic 
in the hard fluctuations (the field $\delta A^\mu$ in  Eq.~(\ref{eq:decomp})). 
The leading order non-trivial contribution  
therefore comes from the cubic terms in the action. (The 
contribution from quartic terms to the evolution is sub--leading in DIS.)

The difference between the two actions is
\be
\Delta S \equiv S_W - S_{\ln W}= {\rm Tr}\left(\rho\, [W-\ln W]\right)
\ee
where $\rho = \rho^a T_a$ 
and $\ln W $ is defined as in Eq.~(\ref{logdef}) to be  
\be
\ln W  \equiv 
\ln [1-(1-W)] = -{\rm Tr}\,\left(\rho\,[(1-W) +{1 \over 2}(1-W)^2 +
 {1\over 3}(1-W)^3 + \cdots]\right)\, .
\ee
The integration over the spatial variables $x^-$ and $x_\perp$ 
and the convolution with $\delta (x^-)$ is implicit in the trace above.
The difference between the two actions is then 
\be
\Delta S = {\rm Tr}\left(\rho\,[{1 \over 2}(1-W)^2 + {1\over 3}(1-W)^3 
+ \cdots]\right)\, ,
\ee
where $1-W$, from Eq.~(\ref{Wilsonline}) can be expanded as 
$1-W= -igA^- + (g^2/2)\,\hat{P} (A^-)^2 + \cdots$.
Again, the integral over $x^+$ is implicit in the expansion, with the 
symbol $\hat{P}$ denoting the time ordering in $x^+$. 
Potential differences between the two actions 
will show up at order $A^2$. At this order~\footnote{We use the 
following conventions for the trace of adjoint matrices: ${\rm Tr}(T^aT^b)
= N_c \delta_{ab}$ and ${\rm Tr}(T^aT^bT^c)= {iN_c\over 2}f_{abc}$.}, 
\be
\Delta S(A^2) \sim {\rm Tr}\left(\,\rho\,(A^-)^2\right) \sim 
\rho_a f_{abc} \int dx^+ dy^+ A^-_b(x^+) A^-_c(y^+).
\ee
This term is identically zero because the integrand is symmetric
under both the color exchange $b \leftrightarrow c$ and the co--ordinate 
exchange $x^+\leftrightarrow y^+$ while   multiplying 
the totally anti--symmetric structure constant $f_{abc}$.

To investigate terms of order $A^3$, it is convenient to
first consider $S_W$ and $S_{\ln W}$ separately. 
The cubic terms in the expansion of  $S_{\ln W}$ are
\be
S_{\ln W}(A^3)&=& {g^3 \over N_c}\,{\rm Tr}\Bigg( 
\rho\, \bigg[\hat{P} (A^-)^3 -
{1 \over 2} A^- \hat{P} (A^-)^2 - {1 \over 2} \hat{P} (A^-)^2 A^-
+ {1 \over 3} (A^-)^3\bigg]\Bigg) \nonumber \\
&=& 
{g^3 \over N_c}  {\rm Tr} \rho \int dx^+ dy^+ dz^+ 
A^-(x^+) A^-(y^+)A^-(z^+) \nonumber \\
&\times&\bigg[\theta (x^+ - y^+) \theta (y^+ - z^+) -{1 \over 2} 
\theta (x^+ - y^+)  -{1 \over 2} \theta (y^+ - z^+) + {1 \over 3}
\bigg] 
\label{eq:lnwc}
\ee
After some algebra (performed in appendix B) the above can be 
re-expressed as 
\be
S_{\ln W}(A^3)&=& 
{g^3 \over 6} \rho_a \int dx^+ dy^+ dz^+ 
A^-_b(x^+) A^-_c(y^+) A^-_d(z^+)
\theta (x^+ - y^+) \theta (y^+ - z^+) 
\nonumber \\
&\times&\bigg[f^{adn}f^{bcn} -f^{abn}f^{cdn}\bigg]
\label{eq:lnwcf}
\ee

The cubic term in $S_W$ is 
\be
S_{W}(A^3)&=&
{g^3 \over N_c} {\rm Tr} \rho \,\hat{P} (A^-)^3\nonumber \\
&=&
{g^3 \over N_c} \rho_a \int dx^+ dy^+ dz^+ 
A^-_b(x^+) A^-_c(y^+) A^-_c(z^+)
\theta (x^+ - y^+) \theta (y^+ - z^+) \nonumber \\
&\times&\bigg[{I_2 \over 6}
(f_{adn}f_{bcn} -f_{abn}f_{cdn}) + d_{abcd}\bigg]\, .
\label{eq:wcubic}
\ee
Here, we have used an identity for the trace of four $SU(3)$ adjoint
matrices~\cite{vermaseren}. For an adjoint representation, $I_2 = N_c$. 
Also, the totally symmetric tensor $d_{abcd}$ is defined as the symmetrized 
trace of four $SU(3)$ adjoint matrices. For an explicit form, see 
Ref.~\cite{Macfarlane}.
Note that the  f-terms above are identical to those
derived from $S_{\ln W}$ in Eq.~(\ref{eq:lnwcf}). However, this 
action also contains the $d_{abcd}$ term that was 
absent in the $S_{\ln W}$ action. 

In appendix C, we 
show that, within the approximations made in the derivation of the 
small $x$ evolution equation in Ref.~\cite{RGE}, the $d_{abcd}$ term 
does not contribute. Therefore $\Delta S = 0$ to cubic order.  One 
may therefore conclude that BFKL equation can 
also be obtained from the $S_{\ln W}$ action.

The reason the $d_{abcd}$ term in the $S_W$ action vanishes is because 
the propagator of the hard modes (and the color sources to which it couples)
is static. The static nature of the sources is due to 
the fact that one ignores the recoil of the color sources as they emit
softer partons. As one goes to a 
next-to-leading-order calculation, one will have to take recoil effects 
into account. These would cause the color sources to be time dependent,
giving rise to a finite contribution from the $d^{abcd}$ terms 
in the $W$ action. Conversely, 
note that the $d^{abcd}$ terms are naturally absent in $\ln W$ action.

The fact that the 
$S_{\ln W}$ action does not have the $d_{abcd}$ 
term suggests that the underlying
symmetry of the small $x$ dynamics is manifest in this action.
The agreement between the two actions is even more remarkable when one 
considers that the factor $1/6$ in Eq.~(\ref{eq:wcubic}) 
comes directly from the trace of 
four adjoint generators, while in the ``$\ln W$'' action it arises as 
a consequence of extensive algebraic manipulations.

\section{Summary}
\vskip 0.1in

In this paper, we proposed an alternative form of the small $x$
effective action to the one discussed in Ref.~\cite{RGE}.  We showed
explicitly that both forms of the effective action are 
compatible with Wong's equations, albeit with currents that satisfy 
different boundary conditions. We showed that the the 
two effective actions agreed up to cubic order in the fields. Consequently, 
both of them give rise to the BFKL equation. However, in the case of 
the effective action of Ref.~\cite{RGE}, one had to explicitly invoke the 
kinematic constraint imposed by the static sources--no such constraint 
was necessary for the action we propose.
Differences between the two actions will show up at higher orders when one 
considers sub--leading corrections to the small $x$ effective action. 

\section*{Acknowledgments}

We would like to thank Alex Kovner, Rob Pisarski and Jens Wirstam 
for reading the
manuscript. We thank Larry McLerran for useful remarks.  One of us (R.V.)
would also like to thank E. Iancu for vigorous discussions and 
(S.J.) would like to thank Harry Lam for insightful comments.  J. J-M. 
would like to thank Alex Kovner for helpful discussions. S. J. was
supported by the Director, Office of Science, Office of High
Energy and Nuclear Physics, Division of Nuclear Physics, and by the
Office of Basic Energy Sciences, Division of Nuclear Sciences, of the
U.S. Department of Energy under Contract No. DE-AC03-76SF00098. J.J-M.
is supported at the University of Arizona by the U.S. Department of
Energy under Contract No. DE-FG03-93ER40792.  R. V. is supported at
BNL by the U.S. Department of Energy under Contract
No. DE-AC02-98CH10886.

\section*{Appendix A}

 The equation of motion for color charges
 in the action (\ref{logwaction}) has already been derived in Section 
 \ref{sec:logwaction}. The derivation is  the same for the point 
 particle action Eq.~(\ref{jsrj1}).
 Further, the Euler-Lagrange equation for the position 
 \be 
 p^I_\mu = m{dx^I_\mu \over d\tau}
 \ee
 follows trivially from Eq.~(\ref{jsrj1}).  
 Hence, we will only derive here the other remaining equation of motion, 
 namely, that for the momentum $p_\mu$.

 The Euler-Lagrange equation for the momentum in Eq.~(\ref{jsrj1}) is 
 \be
 \dot p^I_\mu
 =
 -
 {\delta \over \delta x_I^\mu} S_{\rm Wong}
 \;.
 \ee
 The right hand side is given by 
 \be
 {\delta S_{\rm Wong}\over \delta x_I^\mu(\tau)}
 & = &
 {i\over N_c}\,
 {\delta \over \delta x_I^\mu(\tau)}
 \sum_{J=1}^K
 {\rm Tr}\,
 \left\{
 Q_0^J \ln W_J 
 \right\}
 \non
 & = &
 {i\over N_c}\,
 \int d\tau'\,
 \sum_{J=1}^K
 {\delta A^\nu_a(x_J(\tau')) \over \delta x_I^\mu(\tau)}
 {\delta \over \delta A^\nu_a(x_J(\tau'))}
 {\rm Tr}\,
 \left\{
 Q_0^J \ln W_J 
 \right\}
 \non
 & & {}
 +
 {i\over N_c}\,
 \int d\tau'\,
 \sum_{J=1}^K
 {\delta v_J^\nu(\tau') \over \delta x_I^\mu(\tau)}
 {\delta \over \delta v_J^\nu(\tau')}
 {\rm Tr}\,
 \left\{
 Q_0^J \ln W_J
 \right\}
 \ee
 Using Eqs.(\ref{eq:wc}) and (\ref{eq:Qxplus}),
 we get
 \be
 {\delta S_{\rm Wong}\over \delta x_I^\mu(\tau)}
 & = &
 -
 v^\nu 
 \partial_\mu 
 A_\nu^a(x_I(\tau))\, Q_a^I(\tau) 
 +
 {i\over N_c}\,\int d\tau'\,
 \dot{\delta}(\tau' -\tau)\, 
 {\delta \over \delta v_\mu^I(\tau')}
 {\rm Tr}\,
 \left\{
 Q_0^I \ln W_I 
 \right\}
 \ee
 For simplicity, we'll omit the particle label $I$ from now on. 
 We'll also omit any argument labels 
 (everything should be considered as a function of $\tau$ via $x(\tau)$).
 Applying the method developed in Section \ref{sec:logwaction}, we see that
 \be
 {\delta S_{\rm Wong}\over \delta x^\mu}
 & = &
 -
 Q_a v^\nu \partial_\mu A_\nu^a\, 
 -
 {i\over N_c}
 {d\over d\tau}
 {\rm Tr}\, \left\{ Q_0 {\delta \over \delta v_\mu} \ln W \right\}
 \non
 & = &
 -
 Q_a v^\nu \partial_\mu A_\nu^a\, 
 +
 {d\over d\tau}
 \left\{
 A_\mu^a\, Q_I^a 
 \right\}
 \non
 & = &
 -
 Q_a v^\nu \partial_\mu A_\nu^a\, 
 +
 Q_a v^\nu \partial_\nu A_\mu^a
 +
 A_\mu^a\, 
 {d\over d\tau}
 Q_a 
 \non
 & = &
 -
 Q_a v^\nu \partial_\mu A_\nu^a\, 
 +
 Q_a v^\nu \partial_\nu A_\mu^a
 -
 ig A_\mu^a\, [Q, v^\nu A_\nu]_a
 \non
 & = &
 -Q_a G_{\mu\nu}^a v^\nu
 \ee
 where
 \be
 G_{\mu\nu}^a
 =
 \partial_\mu A_\nu^a\, 
 -
 \partial_\nu A_\mu^a
 +g\,f_{abc}\, A_\mu^b A_\nu^c
 \ee
 This yields the desired result:
 \be
 \dot{p}_\mu = Q_a\, G^a_{\mu\nu} v^\nu  \, .
 \ee

\section*{Appendix B}

We will show here how one gets equation 
(\ref{eq:lnwcf}) from equation (\ref{eq:lnwc}). 
We start with Eq.~(\ref{eq:lnwc}):
\be
S_{\ln W}(A^3)&=& {g^3 \over N_c}\,{\rm Tr}\Bigg(\rho \int dx^+ dy^+ dz^+ 
A^-(x^+) A^-(y^+)A^-(z^+) \nonumber \\
&&\bigg[\theta (x^+ - y^+) \theta (y^+ - z^+) -{1 \over 2} 
\theta (x^+ - y^+)  -{1 \over 2} \theta (y^+ - z^+) + {1 \over 3}
\bigg]\Bigg) \, .\nonumber
\ee
To save space, we will use the following shorthand notation.
We shall represent the Light Cone times $x^+, y^+, z^+$ by $1,2,3$, 
and shall 
not write the Light Cone Lorentz index `$-$' and the integrations over the 
Light Cone times explicitly. For example,
\be
\theta_{123}\equiv \theta (x^+ - y^+)\theta (y^+ - z^+)\, . \nonumber
\ee
We shall
also neglect the overall coefficient $g^3/N_c$, 
including it only at the very last step. With these notations, 
equation (\ref{eq:lnwc}) becomes
\be
S_{\ln W}(A^3)&=& {\rm Tr}\left(
\rho A_1 A_2 A_3 \bigg[\theta_{123} -{1 \over 2} 
\theta_{12}  -{1 \over 2} \theta_{23} + {1 \over 3} \bigg]\right)\, . 
\label{eq:shlnwc}
\ee
We will now use identities like $\theta_{12} +\theta_{21} =1$ to write
Eq.~(\ref{eq:shlnwc}) as
\be
S_{\ln W}(A^3)&=& {\rm Tr}\, \rho A_1 A_2 A_3 \bigg[\theta_{123} -{1 \over 2} 
\theta_{12}(\theta_{23} +\theta_{32})  
-{1 \over 2} \theta_{23}(\theta_{12} + \theta_{21})  \nonumber \\
&+&{1 \over 3}(\theta_{123}+ \theta_{132} +\theta_{231} + \theta_{213}
+ \theta_{312} + \theta_{321})  
\bigg] \nonumber \\
&=&{\rm Tr}\,\rho A_1 A_2 A_3 \bigg[ -{1 \over 2}(\theta_{12}\theta_{32}
+ \theta_{21}\theta_{23}) + {1 \over 3}(\theta_{123}+ \theta_{132} 
+\theta_{231} + \theta_{213} + \theta_{312} + \theta_{321})\bigg]
\nonumber \\
&=&{\rm Tr}\,\rho A_1 A_2 A_3 \bigg[ -{1 \over 2} \theta_{12}\theta_{32}
(\theta_{13} + \theta_{31})  -{1 \over 2} \theta_{21}\theta_{23}
(\theta_{13} + \theta_{31})  \nonumber \\
&+& {1 \over 3}(\theta_{123}+ \theta_{132} 
+\theta_{231} + \theta_{213} + \theta_{312} + \theta_{321})\bigg]
\nonumber \\
&=&{\rm Tr}\,\rho A_1 A_2 A_3 \bigg[ -{1 \over 2} (\theta_{132} +\theta_{312}
+ \theta_{213} +\theta_{231}) \nonumber \\
&+& {1 \over 3}(\theta_{123}+ \theta_{132} 
+\theta_{231} + \theta_{213} + \theta_{312} + \theta_{321})\bigg]
\nonumber \\   
&=&{1\over 6}\,{\rm Tr}\,\rho A_1 A_2 A_3 \bigg[\theta_{123}-  \theta_{213}
 + \theta_{321} -  \theta_{312} +\theta_{123} - \theta_{132} 
+\theta_{321}- \theta_{231}\bigg]\nonumber \\
&=& {1 \over 6}\,{\rm Tr}\,\rho \left( \theta_{123} [A_1,A_2] A_3 
+  \theta_{321} [A_1,A_2],A_3 +  \theta_{123} A_1 [A_2,A_3] 
+  \theta_{321} A_1 [A_2,A_3]\right) \nonumber
\ee 
which, after change of variables, can be re--written as
\be
S_{\ln W}(A^3)&=&{1 \over 6}\, \theta_{123}\,{\rm Tr}\left(\rho 
\left( [[A_1,A_2],A_3]+[[A_3,A_2],A_1]\right)\right) \, .
\ee
Using $[T^a,T^b]=if^{abc} T^c$, and restoring all the indices, 
coefficients, and integration variables, we obtain finally 
\be
S_{\ln W}(A^3)&=&{g^3 \over 6}\, \rho_a\, \int dx^+ dy^+ dz^+ 
\big[A^-_b(x^+) A^-_c(y^+) A^-_d(z^+)
\theta (x^+ - y^+) \theta (y^+ - z^+) \nonumber \\
&\times&\bigg[f_{adn}f_{bcn} -f_{abn}f_{cdn} \bigg]\, ,
\ee    
which is Eq.~(\ref{eq:lnwcf}). 

\section*{Appendix C}

In this appendix, we will show that, within the approximations made in
the Wilson renormalization group approach, the totally symmetric 
$d_{abcd}$--term in Eq.~(\ref{eq:wcubic}) vanishes.
Using the field decomposition Eq.~(\ref{eq:decomp}), we get
\be
S^{sym}_{cubic}&=& {g^3 \over N_c} d^{abcd} \rho_a \int dx^+ dy^+ dz^+ 
\theta (x^+ - y^+) \theta (y^+ - z^+)\nonumber\\ 
&&\bigg[a^-_b(x^+) A^-_c(y^+) A^-_d(z^+) 
+ A^-_b(x^+) a^-_c(y^+) A^-_d(z^+) 
+ A^-_b(x^+) A^-_c(y^+) a^-_d(z^+) \bigg]\, .\nonumber
\ee
Renaming the variables in the second and third terms above gives
\be
S^{sym}_{cubic}&=& {g^3 \over N_c} d^{abcd} \rho_a \int dx^+ dy^+ dz^+ 
a^-_b(x^+) A^-_c(y^+) A^-_d(z^+)\nonumber\\
&&\left[\theta (x^+ - y^+) \theta (y^+ - z^+) + 
\theta (y^+ - x^+) \theta (x^+ - z^+) +
\theta (z^+ - y^+) \theta (y^+ - x^+) \right]. \nonumber
\ee
Re--write the sum of $\theta$--functions above as follows:
\be
&&\theta (x^+ - y^+) \theta (y^+ - z^+) + 
\theta (y^+ - x^+) \theta (x^+ - z^+) +
\theta (z^+ - y^+) \theta (y^+ - x^+) =\nonumber\\
&&\theta (x^+ - y^+) \theta (y^+ - z^+) \theta (x^+ - z^+) + 
\theta (y^+ - x^+) \theta (x^+ - z^+)  \theta (y^+ - z^+) + \nonumber\\
&& \theta (z^+ - y^+) \theta (y^+ - x^+) =\nonumber\\
&&\theta (y^+ - z^+) \theta (x^+ - z^+) +
\theta (z^+ - y^+) \theta (y^+ - x^+) =\nonumber\\
&&\theta (y^+ - z^+) \theta (x^+ - z^+) -
\theta (z^+ - y^+) \theta (x^+ - y^+) + \theta (z^+ - y^+)\nonumber
\ee
The first two terms in the last line are anti--symmetric with respect
to change of $y^+ \rightarrow z^+$, and multiply the product
$d^{abcd} A^-_c(y^+) A^-_d(z^+)$ which is totally symmetric.
They therefore vanish, and the expression in $S_{cubic}^{sym}$ reduces 
to 
\be
S^{sym}_{cubic}&=& {g^3 \over N_c}\,d^{abcd}\,\rho_a \int dx^+ dy^+ dz^+ 
\,\theta (z^+ - y^+) a^-_b(x^+) A^-_c(y^+) A^-_d(z^+) \, .
\label{symcbc}
\ee
This term can further be written as
\be
&&\int dx^+ dy^+ dz^+ 
\theta (z^+ - y^+) a^-_b(x^+) A^-_c(y^+) A^-_d(z^+)=\nonumber\\
&&\int dx^+ dy^+ dz^+ 
\,\bigg[1- \theta (y^+ - z^+)\bigg] a^-_b(x^+) A^-_c(y^+) A^-_d(z^+)\, . 
\ee
Consider the ``$1$'' term on the right hand side:
\be
&&\int dx^+ dy^+ dz^+ 
a^-_b(x^+) A^-_c(y^+) A^-_d(z^+)=
\int dx^+  a^-_b(x^+) dy^+ dz^+ A^-_c(y^+) A^-_d(z^+) 
\sim \nonumber\\
&&\int dx^+  a^-_b(x^+) d(y^+ - z^+) d(y^+ + z^+)  
A^-_c(y^+) A^-_d(z^+) \, .
\ee
When we integrate over hard fluctuations, the term $A^-_c(y^+) A^-_d(z^+)$
will become the hard fluctuations propagator $G^{--}(y^+ - z^+)$. 
After integrating this propagator over the
$d(y^+ - z^+)$ variable, it will give an overall factor $p^-$. 
Since $\rho \sim \delta(p^-)$ (this would break down when considering NLO
corrections), the  ``$1$'' term vanishes
because $p^- \delta (p^-) = 0$. The integrand of Eq.~(\ref{symcbc}) is then 
\be
&&\int dx^+ dy^+ dz^+ 
\theta (z^+ - y^+) a^-_b(x^+) A^-_c(y^+) A^-_d(z^+)=\nonumber\\
&&- \int dx^+ dy^+ dz^+ 
\theta (y^+ - z^+) a^-_b(x^+) A^-_c(y^+) A^-_d(z^+) \, .
\ee
Since the LHS of the above is invariant under 
$z^+\leftrightarrow y^+,\;\;c\leftrightarrow d$, 
the above identity would require it to be `$-$' of itself, and 
therefore equal 
to zero. 
Thus the $d_{abcd}$--term in the cubic piece of the action vanishes,
and we are left with
\be
S_{cubic}&=& {g^3 \over N_c}\,\rho_a \int dx^+ dy^+ dz^+ 
A^-_b(x^+) A^-_c(y^+) A^-_d(z^+)\nonumber\\
&\times&{I_2\over 6}\,\theta (x^+ - y^+) \theta (y^+ - z^+)
\bigg[f^{adn}f^{bcn} -f^{abn}f^{cdn}\bigg] \, ,
\label{eq:expandcubicfterm}
\ee
which is identical to cubic term in the expansion of $S_{\ln W}$.

\end{document}